\documentclass[pra,aps,twocolumn]{revtex4}
\usepackage{graphicx,amsfonts,bm,amsmath}
\newcommand{\bk}{b_{\bm{k}}}
\newcommand{\bkd}{b_{\bm{k}}^{\dag}}
\newcommand{\bmk}{b_{-\bm{k}}}

\newcommand{\sumk}{\sum_{\bm{k}}}
\newcommand{\wk}{\omega_{\bm{k}}}
\newcommand{\fk}{f_{\bm{k}}}
\newcommand{\gk}{g_{\bm{k}}}
\newcommand{\pe}{|1\rangle\!\langle 1|}

\newcommand{\rl}{\rangle\!\langle}
\newcommand{\unit}{\mathbb{I}}

\DeclareMathOperator{\re}{Re}
\DeclareMathOperator{\diag}{diag}

\begin{document}
\title{Complete disentanglement by partial pure dephasing}
\author{Katarzyna Roszak}
 \email{Katarzyna.Roszak@pwr.wroc.pl}
\author{Pawe{\l} Machnikowski}
 \affiliation{Institute of Physics, Wroc{\l}aw University of
Technology, 50-370 Wroc{\l}aw, Poland}

\begin{abstract}
We study the effect of pure dephasing on the entanglement of a pair of
two-level subsystems (qubits). We show that partial dephasing induced by a
super-Ohmic reservoir, corresponding to well-established properties of
confined charge states and phonons in semiconductors,
may lead to complete disentanglement. We show also that the 
disentanglement effect increases with growing distance between the
two subsystems. 
\end{abstract}


\maketitle

\section{Introduction}

Entanglement \cite{einstein35,schrodinger35b} is one
of the fundamental constituents of quantum theory. 
Correlations between the results of appropriately chosen measurements
on entangled subsystems cannot be accounted for by any classical
(realistic and local) theory \cite{bell64}, precluding the existence
of a wide class of hypothetical more fundamental structures underlying the
incompleteness of quantum description. Apart from its essential
role in our understanding of the quantum world, entanglement is an
important resource in quantum information processing \cite{nielsen00}
where it provides a quantum channel for teleportation
\cite{bennett93}, superdense coding \cite{bennett92}, and
distribution of cryptographic keys \cite{ekert91}.

In order to manifest genuinely quantum behavior resulting from
entanglement a quantum system must maintain phase relations
between the components of its quantum superposition state, involving
different states of distinct subsystems. Keeping in mind that the
subsystems may be separated by a macroscopic distance,
one may expect such a non-local superposition state to be extremely
fragile to the dephasing effect of the environment. In fact, it has
been shown that entanglement of a pair of two-level
subsystems tends to decay faster than local coherence
\cite{yu02,yu03,yu04}. As expected, the decay of entanglement is
stronger if the subsystems interact with different environments (which
might result from a large spatial separation between them): certain
states that show robust entanglement under collective dephasing become
disentangled by separate environments
\cite{yu03}. Remarkably, it was shown for different classes
of systems \cite{diosi03,yu04,dodd04} that certain states may become
separable 
(completely disentangled) within a final time under conditions that
lead to usual, exponential decay of local coherence. Since even
partial entanglement of many copies of a bipartite quantum system may
still be distilled to a smaller number of maximally entangled systems
\cite{horodecki97}, many quantum information processing and
communication tasks may be carried out using  partly disentangled
systems, as long as there is still some entanglement left. 
It is therefore essential to understand whether
environmental influence leads to the appearance of separability
in realistic models of dephasing.

In this paper we study the decay of entanglement between a pair of
spatially localized two-level systems (qubits) under pure dephasing
induced by a bosonic bath. In order to capture the physical aspects of
disentanglement in real systems, we focus on the specific case of two
excitons confined in one or two semiconductor quantum dots and coupled
to phonons. In this
system, the properties of the coupling and the resulting kinetics of
dephasing are well understood. In particular, the experimentally
observed evolution of the system may be accounted for by pure
dephasing within an independent boson model \cite{vagov04}. In
contrast to the properties of the Ohmic model \cite{breuer02}, 
the carrier-phonon dynamics in a real system leads only
to \textit{partial} 
dephasing \cite{borri01}: non-diagonal elements of
the density matrix do not vanish exponentially but rather
decay only to a certain finite value (apart from
other processes on much longer time scales). 
This behavior is characteristic of super-Ohmic 
spectral densities \cite{krummheuer02,alicki04b} resulting from
the actual carrier-phonon couplings and phonon density of states 
\cite{mahan00}. The effect of pure dephasing on a system of two
excitons in a single quantum dot was studied in a recent work
\cite{axt05b} in the context of ultrafast optical excitation. 
Here we describe the dynamics of two spatially separated systems,
focusing on the resulting loss of entanglement. 

The main result of this paper is that a physical mechanism, based on a
microscopic model of interactions between charges and phonons in solids
and quantitatively confirmed by experiments on real quantum dot
systems, may lead to the \textit{complete decay} 
of the entanglement of a pair of two-level
systems. This is in striking contrast with the partial
asymptotic decay of local
coherence under the same environmental dephasing.
Moreover, using the physical
coupling constants for spatially localized states we are able to
describe the effect of spatial separation on the evolution of
entanglement and to physically account for the crossover from a collective
to an individual reservoir regime. 
These results are of importance not only for the general understanding
of the properties of entanglement of open systems but also for
practical tasks related to coherent control of charge states in
multi-partite semiconductor systems (like carrier states in coupled quantum
dots). This, in turn, may affect the feasibility of solid-state based
quantum computing schemes \cite{biolatti00,calarco03}, including
quantum error correction based on collective encoding
of logical qubits (concatenation) \cite{knill05}.

The paper is organized as follows. In Sec.~\ref{sec:model} we define
the model of the excitonic two-qubit system and find its
evolution. Next, in Sec.~\ref{sec:disent}, we discuss the evolution of
entanglement between the two qubits. Sec.~\ref{sec:concl} concludes
the paper with final remarks.

\section{The model and evolution of the two-qubit system}
\label{sec:model}

In this Section we describe the evolution of the two excitonic qubits under
the dephasing effect of phonons. The result will be used in the next
Section for the calculation of entanglement between the two qubits.

The system is described by the Hamiltonian
\begin{eqnarray}\label{ham0}
H & = & \epsilon_{1}(\pe\otimes\unit)
+\epsilon_{2}(\unit\otimes\pe)
+\Delta\epsilon(\pe\otimes\pe) \nonumber \\ 
&&+(\pe\otimes\unit)\sumk\fk^{(1)}(\bkd+\bmk) \\ \nonumber
&&+(\unit\otimes\pe)\sumk\fk^{(2)}(\bkd+\bmk)
+\sumk\wk\bkd\bk ,
\end{eqnarray}
where the two states of each subsystem are denoted by
$|0\rangle$ and $|1\rangle$, $\mathbb{I}$ is the unit operator,
$\epsilon_{1,2}$ are the transition energies
in the two subsystems, $\Delta\epsilon$ is an energy shift
due to the interaction between the subsystems, $\fk^{(1,2)}$
are system-reservoir coupling constants, $\bk,\bkd$ are bosonic
operators of the reservoir modes, and $\wk$ are the corresponding energies
(we put $\hbar=1$). The explicit tensor notation refers to the two
subsystems but is suppressed for the reservoir components.

Exciton wave functions will be modelled by anisotropic Gaussians
with the extension $l_{\mathrm{e/h}}$ in the $xy$ plane
for the electron/hole and $l_{z}$
along $z$ for both particles.
Then, the coupling constants for the deformation potential coupling between
confined charges and longitudinal phonon modes have the form 
$\fk^{(1,2)}=\fk e^{\pm ik_{z}d/2}$, where $d$ is
the distance between the subsystems ($d=0$ corresponds to a biexciton,
i.e., two excitons with different spins in a single QD) and 
$\fk=\fk^{(\mathrm{e})}-\fk^{(\mathrm{h})}$, with 
\begin{displaymath}
\fk^{(\mathrm{e/h})}=\sigma_{\mathrm{e/h}}\sqrt{\frac{k}{2\varrho Vc}}
\exp\left[
-\frac{l_{z}^{2}k_{z}^{2}+l_{\mathrm{e/h}}^{2}k_{\bot}^{2}}{4}\right].
\end{displaymath}
Here $V$ is the normalization volume of the bosonic reservoir,  
$k_{\bot/z}$ are momentum
components in the $xy$ plane and along the $z$ axis,
$\sigma_{\mathrm{e/h}}$ are deformation potential constants for
electrons/holes, $c$ is the speed of longitudinal sound,
and $\varrho$ is the crystal density. 
In our calculations we put $\sigma_{\mathrm{e}}=8$ eV,
$\sigma_{\mathrm{h}}=-1$ eV, $c=5.6$ nm/ps, $\varrho=5600$ kg/m$^{3}$
(corresponding to GaAs), and
$l_{\mathrm{e}}=4.4$ nm, $l_{\mathrm{h}}=3.6$ nm, $l_{z}=1$ nm. 

Although the above coupling constants correspond to a specific system,
their form reflects rather general physical 
conditions. First, for any translationally invariant reservoir, 
$\bm{k}$ may be interpreted as
momentum. Second, the system is assumed to be localized in space and
to interact with the reservoir via a spatially local interaction (this
is natural for solid state systems and corresponds to the dipole
approximation of electrodynamics). Then, if the system size is $l$
the uncertainty of momentum is $\sim 1/l$ and momentum
conservation allows coupling to bosonic modes only within this range
of $\bm{k}$, as manifested by the Gaussian cutoff in our formula. The
position-dependent phase is also a general feature, while the
analytical behavior at $k\to 0$ is specific to the coupling. 

The Hamiltonian (\ref{ham0}) is diagonalized by the
transformation $\mathbb{W}=\sum_{i=1}^{3}|i\rl i|W_{i}$, where we
use the standard product basis $|0\rangle\equiv|0\rangle|0\rangle$,
$|1\rangle\equiv|0\rangle|1\rangle$, etc., and $W_{i}$ are Weyl shift
operators 
\begin{displaymath}
W_{i}=\exp[\sumk\gk^{(i)*}\bk-\mathrm{H.c.}],
\end{displaymath}
where $\gk^{(1,2)}=\fk^{(1,2)}/\wk$ and $\gk^{(3)}=\gk^{(1)}-\gk^{(2)}$. 
Upon this unitary transformation one gets
\begin{displaymath}
\tilde{H}=\mathbb{W}H\mathbb{W^{\dag}}
=\tilde{H}_{\mathrm{L}}+\tilde{H}_{\mathrm{I}}+\tilde{H}_{\mathrm{res}},
\end{displaymath}
where 
\begin{displaymath}
\tilde{H}_{\mathrm{L}}=E_{1}(\pe\otimes\unit)+E_{2}(\unit\otimes\pe)
\end{displaymath}
describes the independent (local) evolution
of the subsystems, 
\begin{displaymath}
\tilde{H}_{\mathrm{I}}=\Delta E(\pe\otimes\pe)
\end{displaymath}
describes their interaction, and 
\begin{displaymath}
\tilde{H}_{\mathrm{res}}=\sumk\wk\bkd\bk
\end{displaymath}
is the reservoir Hamiltonian. The energies here are 
\begin{displaymath}
E_{i}=\epsilon_{i}-\sumk\wk|\gk^{(i)}|^{2}
\end{displaymath}
and 
\begin{displaymath}
\Delta E=\Delta\epsilon-2\re\sumk\wk\gk^{(1)}\gk^{(2)*}. 
\end{displaymath}
The evolution operator generated by the Hamiltonian (\ref{ham0}) may
now be written as
$U_{t}=\mathbb{W}^{\dag}\mathbb{W}_{t}\tilde{U}_{t}$, where 
$\tilde{U}_{t}=\exp(-i\tilde{H}t)$ and 
$\mathbb{W}_{t}=\tilde{U}_{t}\mathbb{W}\tilde{U}_{t}^{\dag}$. Since
$\tilde{H}$ is diagonal, the evolution described by $\tilde{U}_{t}$ is
trivial. The final formulas are further simplified by performing the
local unitary rotation 
$U_{\mathrm{L}}=\exp(i\tilde{H}_{\mathrm{L}}t)$
(note that $[U_{\mathrm{L}},\mathbb{W}_{t}]=0$).
Then, for the reduced density matrix
of the two-qubit system one finds 
$\rho(t)=U_{\mathrm{L}}^{\dag}\tilde{\rho}(t)U_{\mathrm{L}}$, with
\begin{equation}\label{ro}
\tilde{\rho}(t)
= \mathrm{Tr_{R}}\left[ 
 \mathbb{W}^{\dag}\mathbb{W}_{t}e^{-i\tilde{H}_{I}t}
(\rho_{0}\otimes\rho_{T})e^{i\tilde{H}_{I}t}\mathbb{W}^{\dag}_{t}\mathbb{W}
\right],
\end{equation}
where we assumed that the system is initially in a product state
with the reservoir in the thermal equilibrium state $\rho_{T}$.
Since local unitary transformations do not change the amount of
entanglement, we may use the density matrix $\tilde{\rho}(t)$ instead
of $\rho(t)$ in the calculations.

The elements of the density matrix $\tilde{\rho}(t)$ are found
using Eq.~(\ref{ro}), the definition of the operator $\mathbb{W}$, 
and rules for multiplying and averaging Weyl operators 
\cite{mahan00,roszak05a}.
The result may be written in the operator sum representation
\begin{equation}\label{state}
\tilde{\rho}(t)=\sum_{\mu}UK_{\mu}\tilde{\rho}(0)K^{\dag}_{\mu}U^{\dag}
\end{equation}
with the unitary operator
\begin{eqnarray*}
U&=&|0\rl 0|+\exp\left[i\sumk |\gk|^{2}\sin\wk t\right]
\left( |1\rl 1|+|2\rl 2| \right)  \\
&&
+\exp\left[  4i\sumk |\gk|^{2}\cos^{2}\frac{k_{z}d}{2}\sin\wk t-i\Delta Et
\right] |3\rl 3|
\end{eqnarray*}
and the set of Kraus operators
\begin{eqnarray*}
K_{0} & = & \diag[a(t),b(t),b(t),a(t)], \\
K_{1} & = & \diag[(a^{2}(t)-1)\sqrt{a^{2}(t)+1},0,0,0],\\
K_{2} & = & \sqrt{1-a^{2}(t)}\diag[-a^{2}(t),0,0,1], \\
K_{3} & = & \diag[0,0,(b^{2}(t)-1)\sqrt{b^{2}(t)+1},0],\\
K_{4} & = & \sqrt{1-b^{2}(t)}\diag[0,1,-b^{2}(t),0], 
\end{eqnarray*}
where 
\begin{subequations}
\begin{eqnarray}\label{ab}
a(t) & = & \exp\left[ 
\sumk|\gk|^{2}\cos^{2}\frac{k_{z}d}{2} 
(\cos\wk t-1)(2n_{\bm{k}}+1) \right] ,\\
b(t) & = & \exp\left[  
\sumk|\gk|^{2}\sin^{2}\frac{k_{z}d}{2}
(\cos\wk t-1)(2n_{\bm{k}}+1) \right] ,
\end{eqnarray}
\end{subequations}
and $\gk=\fk/\wk$.
The set of operators $K_{\mu}$ corresponds to a kind of two-qubit phase
damping channel. This channel is generated by the same physical
pure dephasing process that would lead to the usual
phase-damping channel for a single qubit \cite{preskill}. The
operators $K_{1,2}$ may be interpreted as an effect of a ``charge
detector'' which is sensitive only to the total number of excitons in
the system. On the other hand, $K_{3,4}$ reflect the action of a
``discriminator'' which detects in which of the two dots the exciton
is present. Since the two subsystems are identical they may be
distinguished by the reservoir only because of their different position.
Therefore, the latter contribution to dephasing is inefficient for
$d=0$. For a system of two excitons confined in a single dot ($d=0$) 
the present result reduces to that found in Ref. \cite{axt05b}.

\begin{figure}[tb] 
\begin{center} 
\unitlength 1mm
\begin{picture}(80,25)(0,6)
\put(0,0){\resizebox{80mm}{!}{\includegraphics{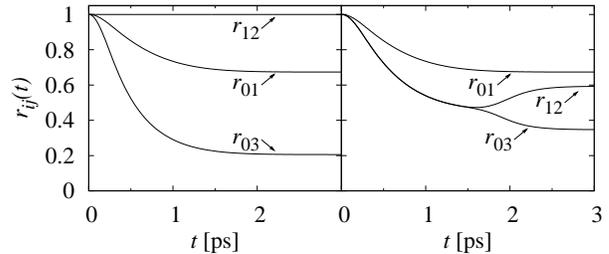}}}
\end{picture} 
\end{center} 
\caption{\label{fig:elem}
The evolution of a two-qubit state
under phonon-induced pure dephasing. The plot shows the relative
reduction of the non-diagonal elements of the reduced density matrix,
$r_{ij}=|\rho_{ij}(t)/\rho_{ij}(0)|$ for
$T=40$ K, $d=0$ (left), and $d=6$ nm (right). For identical qubits one
has $r_{01}=r_{02}=r_{13}=r_{23}$.}
\end{figure}

The dephasing action of the reservoir develops in the course of the
joint carrier-phonon evolution. Initially, all the Kraus operators
except for $K_{0}$ are null.
For long times, the factors $\cos\wk t$ in Eqs. (\ref{ab},b) 
become quickly oscillating
functions of $\bm{k}$ and their contribution averages to 0. 
Consequently, the operators $K_{\mu}$ reach a certain asymptotic
form. As a result, the non-diagonal elements of the density matrix 
decrease form their initial
value to a certain asymptotic value depending on material
parameters, system geometry and temperature (see
Fig.~\ref{fig:elem}). For the
actual carrier-phonon couplings in a semiconductor system 
this asymptotic value is \textit{nonzero}, 
similarly as in the single-qubit dephasing \cite{krummheuer02} 
and the phase information is only partly erased. 
The non-monotonous evolution
of coherence is due to the fact that the process is non-Markovian
and some coherence may be regained in the course of system-reservoir
interaction before the process is completed. 
It is also clear from
Fig.~\ref{fig:elem} that the coherence between the states 
$|1\rangle\equiv|0\rangle|1\rangle$ and
$|2\rangle\equiv|1\rangle|0\rangle$ may be broken only if the
reservoir can distinguish between the two systems (i.e., for $d\neq 0$).

In order to interpret the time dependence of the matrix elements for
$d\neq 0$, shown in Fig.~\ref{fig:elem} (right), let us recall that
the physical mechanism of dephasing is the emission of phonons
from the excitonic qubit to the bulk of the crystal
\cite{jacak03b,vagov02a}. These phonons
form a spherical wave packet around the quantum dot, 
extending with the speed of sound and thus carrying the information
from the qubit to the outside world \cite{roszak05a}.
For short times $t\lesssim d/c$ this reservoir
perturbation pertaining to each qubit stays localized around this
qubit (the phonon wave packets do not overlap) 
so that detecting the presence of a charge is as efficient as
deciding in which dot it is localized. As a result, the coherences
between the ground and two-exciton state (matrix element $\rho_{03}$)
and between the two single-exciton states ($\rho_{12}$) are affected
to the same extent. For longer times, the wave packets
originating from the two qubits intersect which, on one hand, leads
to positive interference and to increased efficiency of overall charge
detection. This is reflected in the right panel of Fig.~\ref{fig:elem} 
by a step-wise drop of $\rho_{03}$ at $t\sim d/c$. 
On the other hand, however, this partial overlap makes it
harder to identify the origin of the perturbation, hence the
``discriminator'' measurement becomes less efficient. Therefore, at
$t\gtrsim d/c$ the inter-qubit coherence $\rho_{12}$ becomes less affected.

\section{Disentanglement by pure dephasing}
\label{sec:disent}

For a quantitative description of the decay of entanglement, a measure
of entanglement that may be calculated from the system state is
needed. For pure states, the von Neumann entropy of one subsystem
\cite{bennett96b} may be used but for mixed states
there is no unique entanglement measure
\cite{bennett96a,bennett96}. 
In order to quantitatively describe entanglement of systems in mixed
states one defines various quantities, which may give essentially
different entanglement measures. Nonetheless, any such quantity must 
satisfy certain natural requirements in order to be useful as 
an entanglement measure: it must vanish for separable states and be
non-zero for entangled ones; it must reduce to the von Neumann entropy
for a pure state; it must be non-increasing under local operations (it
is impossible to increase entanglement by manipulating only one subsystem).
One choice is to use
the \textit{entanglement of formation} (EOF), defined as 
the ensemble
average of the von Neumann entropy minimized over all ensemble
preparations of the state \cite{bennett96a,bennett96}, and thus being
a natural generalization of the von Neumann entropy to mixed states. 
Such a measure may be interpreted as the
asymptotic number of pure singlets necessary to prepare the state by
local operations and classical communication. A practical characterization for
mixed state entanglement is available for small systems
\cite{peres96a,horodecki96} but an explicit formula for
calculating an entanglement measure is known only for 
the EOF of a pair of two-level systems \cite{hill97,wootters98}.  

We will perform the calculations for two initial pure states  
\begin{subequations}
\begin{eqnarray}\label{ini1}
|\psi_{0}^{(1)}\rangle & = & 
\frac{|00\rangle+|01\rangle+|10\rangle-|11\rangle}{2}, \\
\label{ini2}
|\psi_{0}^{(2)}\rangle & = & 
\frac{|01\rangle-|10\rangle}{\sqrt{2}}.
\end{eqnarray}
\end{subequations}
It is assumed that these states are prepared within a time shorter than
reservoir response times.
Both initial states (\ref{ini1}) and (\ref{ini2}) are maximally
entangled, i.e. their entanglement of formation is equal to 1.

Using Eq.~(\ref{state}), we calculate the EOF of the
system state at a time $t$ 
from the Wootters formula \cite{hill97,wootters98} 
\begin{displaymath}
E[\rho(t)]=E[\tilde{\rho}(t)]=-x_{+}\log_{2}x_{+}-x_{-}\log_{2}x_{-},
\end{displaymath}
where
\begin{displaymath}
x_{\pm}=\frac{1\pm\sqrt{1-C^{2}[\tilde{\rho}(t)]}}{2}
\end{displaymath}
and $C[\tilde{\rho}(t)]$ is the
concurrence, given by 
\begin{equation}\label{concurr}
C[\tilde{\rho}(t)]=\max(0,\lambda_{0}-\lambda_{1}-\lambda_{2}-\lambda_{3}),
\end{equation}
where $\lambda_{i}$ are the eigenvalues of the matrix 
$\tilde{\rho}(t)(\sigma_{y}\otimes\sigma_{y})
\tilde{\rho}(t)(\sigma_{y}\otimes\sigma_{y})$ in decreasing order. 

\begin{figure}[tb] 
\begin{center} 
\unitlength 1mm
\begin{picture}(80,53)(0,6)
\put(0,0){\resizebox{80mm}{!}{\includegraphics{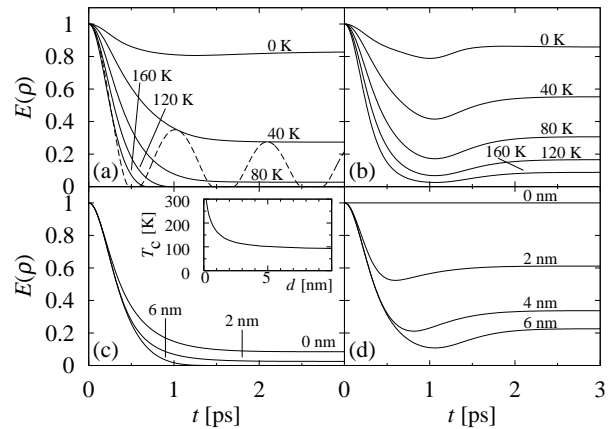}}}
\end{picture} 
\end{center} 
\caption{\label{fig:evol}
Evolution of entanglement of the two-qubit system at various
temperatures with $d=6$ nm (a,b) and for various distances $d$ at $T=100$ K
(c,d). The left panels (a,c) show the result for the initial 
state~(\ref{ini1}) and the right ones (b,d) for the singlet state
(\ref{ini2}). The inset in (c) shows the lowest temperature at which
the decay becomes complete for a given distance. Solid lines
correspond to $\Delta E=0$ and the dashed line to 
$\Delta E/\hbar=6$ ps$^{-1}$ (at $T=40$ K).}
\end{figure}

The evolution of the EOF of the qubit pair is shown in
Fig.~\ref{fig:evol}. In the absence of energy shift $\Delta E$ (solid
lines),
entanglement decays on a time scale
of a few picoseconds. At low temperatures or for
overlapping systems, this process resembles the decay of 
coherences shown in Fig.~\ref{fig:elem}. 
However, for a sufficiently large separation
between the systems and at sufficiently high temperatures the
initially maximal entanglement present in the state (\ref{ini1})
decays completely after a finite time even
though the environment-induced dephasing is always only partial (see
Fig.~\ref{fig:evol}a,c). 
The temperature $T_{\mathrm{c}}$ at which the system becomes separable
is related to the distance as shown in the inset in
Fig.~\ref{fig:evol}c. 
On the other hand, for the other initial state
[Eq.~(\ref{ini2})], the destruction of entanglement is always only
partial
(Fig.~\ref{fig:evol}b,d).

An important case is that of $\Delta E\neq 0$. Such an energy shift
(known as the biexcitonic shift in a semiconductor system) leads to
an entanglement-generating evolution. This mechanism is used 
for performing nontrivial two-qubit gates (controlled-shift) in many
proposals for semiconductor-based quantum information processing
\cite{biolatti00,pazy03a}. As can be seen in Fig.~\ref{fig:evol}a
(dashed line), in the presence of phonon-induced pure dephasing the
cyclic evolution of entanglement is damped and the maximum achievable 
level of
entanglement is reduced. Moreover, extended periods of time appear when
the entanglement remains zero.

The appearance of complete disentanglement for some initial states
under sufficiently strong partial pure dephasing may be understood
with the help of Eq.~(\ref{concurr}). If the completely dephased state
(with a diagonal density matrix)
has $\lambda_{0}-\lambda_{1}-\lambda_{2}-\lambda_{3}<0$ then, by
continuity, it will be surrounded by states with vanishing
concurrence, thus separable.
In this case entanglement vanishes for sufficiently strongly dephased
states, before the complete dephasing is reached. 
For a diagonal density matrix one finds 
$\lambda_{0}-\lambda_{1}-\lambda_{2}-\lambda_{3}=
-2\min(\rho_{00}\rho_{33},\rho_{11}\rho_{22})$, so that the above
condition may only be satisfied if all four diagonal elements are
nonzero. For instance, the totally mixed state (with a density
matrix proportional to unity) is surrounded by a ball of separable
states \cite{zyczkowski}. 
Out of the initial states considered here, the first one 
[Eq.~(\ref{ini1})] satisfies the above condition (it decays towards the
totally mixed state) but the second one [Eq.~(\ref{ini2})] does not.
In general,
the structure of entangled and separable states around the
final state may be quite complicated. 

\begin{figure}[tb] 
\begin{center} 
\unitlength 1mm
\begin{picture}(56,25)(0,6)
\put(0,0){\resizebox{56mm}{!}{\includegraphics{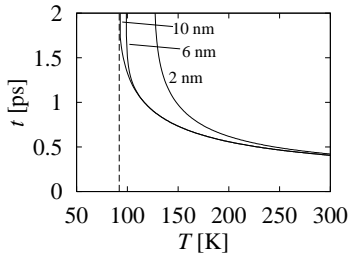}}}
\end{picture} 
\end{center} 
\caption{\label{fig:czas-temp}
The time at which the entanglement of the initial state (\ref{ini1}) decays
completely as a function of temperature.}
\end{figure}

The time at which the entanglement of the state (\ref{ini1}) 
vanishes completely depends on 
temperature and on the distance between the systems. As can be seen in
Fig.~\ref{fig:czas-temp}, this time becomes finite only at a certain
temperature (dashed line in the figure). Slightly above this
critical temperature, complete
disentanglement takes place only for strongly separated systems. For
higher temperatures the disentanglement time for non-overlapping
systems depends very weakly on the distance.
It should be stressed that the appearance of complete
disentanglement at increased temperatures is only related to stronger
dephasing and, in principle, the state might become separable
already at $T=0$ if the coupling were sufficiently strong. 

\begin{figure}[tb] 
\begin{center} 
\unitlength 1mm
\begin{picture}(80,25)(0,6)
\put(0,0){\resizebox{80mm}{!}{\includegraphics{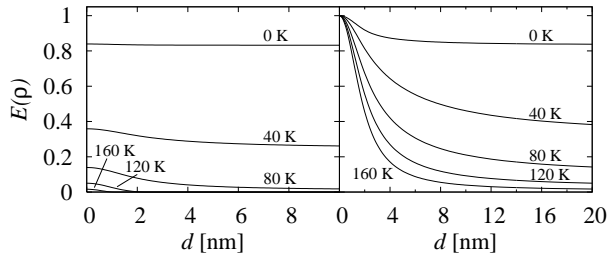}}}
\end{picture} 
\end{center} 
\caption{\label{fig:conc-d}
The asymptotic value of the EOF as a function of
the distance between the subsystems. Left and 
right panels correspond to the initial states (\ref{ini1}) and
(\ref{ini2}), respectively.} 
\end{figure}

On the other hand, the dependence on the separation between the two
subsystems 
reflects a more fundamental crossover, from the regime of a common
reservoir to that of independent reservoirs \cite{yu03}. In our model,
the non-diagonal element $\rho_{12}$ is unaffected for completely
overlapping systems ($d=0$) (see Fig.~\ref{fig:elem}). 
This single non-vanishing element is enough to prevent
complete disentanglement at any temperature and for any coupling
strength. Only when the distance between the systems
grows, dephasing is able to decrease this element to the extent sufficient
for the total destruction of entanglement. This distance effect is shown
in Fig.~\ref{fig:conc-d}, where we plot the asymptotic (long-time)
value of the EOF. 

The effect of spatial separation between the systems is very
clearly visible for the second initial state [Eq.~(\ref{ini2})], which
involves \textit{only} this robust non-diagonal element (see
Fig.~\ref{fig:conc-d}, right). Here, the entanglement is absolutely stable if
the systems overlap (see also Fig.~\ref{fig:evol}d) 
but becomes much more fragile as soon as the
separation between the systems is comparable to their size. This
demonstrates that the distance between the subsystems is the physical
parameter that governs the crossover between the two regimes of
entanglement decay.

\section{Conclusions}
\label{sec:concl}

We have studied the effect of pure dephasing on the
entanglement of two-level subsystems. We have shown
that partial dephasing
may be sufficient to completely destroy entanglement for a class of
initial states 
that may be easily characterized. Apart from its
dependence on the initial state and temperature, the disentanglement
effect shows essential dependence on the distance between the
subsystems, manifesting a crossover between two regimes of
reservoir-induced dephasing. 
Complete disentanglement appears only for spatially separated systems.

Complete disentanglement
due to partial pure dephasing typical for localized carrier states in
semiconductor systems is not only of general interest but also of
relevance to solid state implementations of quantum information
processing. Moreover, understanding the role of the distance
between subsystems in maintaining quantum correlations between them
seems to be essential for realistic design of quantum error correction
schemes based on collective encoding of logical qubits
(concatenation).

One should note that, due to a purely non-Markovian character of the
dephasing effect, the system evolution depends on the way in which the
``initial'' state has been prepared from the ground system
state. Therefore, a reduction of the destructive effect may be
expected if the preparation is done either slowly (adiabatically) or by
shaped pulses \cite{alicki04a,axt05a}. 

Supported by the Polish MNI under 
Grants No. PBZ-MIN-008/P03/2003 and 2 PO3B 085 25. 
P.M. is grateful to A. von Humboldt Foundation for support.


\end{document}